\documentclass[english,aps,pra,reprint,noshowpacs,superscriptaddress]{revtex4-1}

\usepackage[T1]{fontenc}	
\usepackage[latin9]{inputenc}	

\usepackage{geometry}		
\usepackage{graphicx}
\usepackage{amsmath}
\usepackage{bm}
\usepackage{latexsym}
\usepackage{float}
\usepackage{gensymb}
\usepackage{amssymb}
\usepackage{verbatim}
\usepackage{array}
\usepackage{multirow}
\usepackage{listings}
\usepackage{courier}
\usepackage[normalem]{ulem}
\usepackage{url}
\usepackage{tikz}

\newcommand{\compressedaffil}[1]{\affiliation{\footnotesize #1}}

\definecolor{magenta}{rgb}{1.0,0.0,1.0}
\definecolor{indigo}{RGB}{75,0,130}
\definecolor{OliveGreen}{RGB}{85,107,47}

\begin{document}

\title{Studying community development: a network analytical approach}
\author{C. A. Hass}
\email{chris.hass@shaw.ca}
\compressedaffil{Department of Physics, Kansas State University, 1228 N. 17th St., Manhattan, KS 66506, USA}
\author{Florian Genz}
\compressedaffil{ZuS - Future Strategy of Teacher Education, University of Cologne, Germany}
\author{Mary Bridget Kustusch}
\compressedaffil{Department of Physics, DePaul University, 2219 North Kenmore Avenue Suite 211, Chicago, IL 60614, USA}
\author{Pierre-P.~A.~Ouimet}
\compressedaffil{Department of Physics, University of Regina, Regina SK S4S 0A2 Canada}
\author{Katarzyna Pomian}
\compressedaffil{Department of Physics, DePaul University, 2219 North Kenmore Avenue Suite 211, Chicago, IL 60614, USA}
\author{Eleanor C. Sayre}
\compressedaffil{Department of Physics, Kansas State University, 1228 N. 17th St., Manhattan, KS 66506, USA}
\author{Justyna P. Zwolak}
\compressedaffil{Joint Center for Quantum Information and Computer Science, University of Maryland, College Park, MD 20742, USA}

\begin{abstract}
Research shows that community plays a central role in learning, and strong community engages students and aids in student persistence. Thus, understanding the function and structure of communities in learning environments is essential to education. We use social network analysis to explore the community integration of students in a pre-matriculation, two-week summer program. Unlike previous network analysis studies in PER, we build our networks from classroom video that has been coded for student interactions using labeled, directed ties.  We also examine the change in student conversation topicality over the course of the program, and its connection to the forming student collaborations. We define 3 types of interaction: on-task interactions (regarding the assigned task), on-topic interactions (having to do with science, technology, engineering, and mathematics (STEM)), and off-topic interactions (unrelated to the assignment or STEM). While we do not see a significant change in network analysis measures, we do find fewer off-task interactions later in the program, suggesting that the need for these interactions to negotiate the collaboration is reduced.
\end{abstract}

\maketitle

\section{Introduction}

    Strong community participation increases student persistence and success in university courses~\cite{Rogoff1996}. For instance, recent studies show that students who become well integrated in the in- and out-of-class networks are much more likely to persist in the introductory physics courses than those who do not establish such networks~\cite{Zwolak17-IIP,ZwolakPers2018}. To increase student community integration and  educational success we must understand how student communities evolve. Building on research by Pomian~{\it et al.}~\cite{Pomian2017}, we attempt to use social network analysis (SNA) to observe changes in the focus of student conversations during classroom group work sessions. 
    
    The evolution of a network can be best observed in a closed environment (i.e., with each person's data accessible) where individuals have plenty of opportunities to interact with one another. We use the Integrating Metacognitive Practices and Research to Ensure Student Success (IMPRESS) program -- a summer program for incoming undergraduate students at Rochester Institute of Technology (RIT) -- as our case study.
    
    Social networks represent the structure of a social group and interactions between members of this group. They have been used recently for quantitative analyses of, among other things, student retention, student self-efficacy, and the coauthorship network~\cite{Brewe2012Networks,Zwolak17-IIP,ZwolakPers2018,Dou2016,Bruun2013,Anderson2017}. Most commonly, social networks are generated from surveys or electronic repositories, such as email exchange, Twitter, or Facebook data. In this paper, we use social networks to measure the integration of the IMPRESS~\cite{Franklin2017IMPRESS} student community. In particular, we combine longitudinal video observations with social network analysis to understand how the communities change throughout the program by comparing data from early and late in the program.
    
    While we don't see a change in network measures, we also examine the evolution of student conversation topicality over the course of the program. Our data show that early in this program, student conversation topic during classroom activities often strays from the task at hand. As the program persists, classroom topicality becomes almost solely focused on the task at hand. Building on work done by Langer-Osuna~{\it et al.}~\cite{Langer-Osuna2018} on the role of off-task interactions, we suggest that over the course of the program students build collaborations, an important step in building a community of practice~\cite{Wenger1998}.

\section{Learning Communities}

    Community forms an important part of learning in a post-secondary environment~\cite{Tinto97-CAC}. Being part of a student community allows you to get help with assignments, exam preparation, course selection, and presentations. It also provides emotional support and opportunity for socialization. More formally, 
    in his work on communities of practice, Wenger formalizes the key role that communities play in facilitating learning~\cite{Wenger1998}. Learning occurs through participation in legitimate community practice.
 
    The IMPRESS program allows students to engage in some legitimate activities of a scientific community of practice, blended with a learning community~\cite{Franklin2017IMPRESS,Irving2014AdLab}. Understanding how these communities form and operate is an important part of improving physics education.
    
    In their work, Langer-Osuna {\it et al.} explore the role of non-task oriented conversation in learning~\cite{Langer-Osuna2018}. They observe that off-task conversation serves as a tool for students to build collaborations, and position themselves within collaborations. As students' roles within the collaboration 
    become established, off-task conversation becomes less necessary.
    
    The frameworks provided by Wenger~\cite{Wenger1998}, Langer-Osuna~{\it et al.}~\cite{Langer-Osuna2018}, and Irving and Sayre~\cite{Irving2014AdLab} provide a basis for us to understand the shifts in conversation topicality in the IMPRESS classroom discussions. 
    We explore two possible reasons for major topical shifts in classroom discussion. The nature of different activities could drive different kinds of discussion. In this case, we would likely see different conversation topicality distribution in different activities or types of activity. A second possibility is that conversation topicality is driven by formation of a successful collaboration. In this case, we would see topicality distribution change from early in the program to late in the program, as student membership in the collaboration shifts. Days at similar times in the program would display similar topicality distribution.
    
\section{What is IMPRESS?}\label{sec:context}

\begin{figure}[b]
 \includegraphics[width=0.35\textwidth, angle=270]{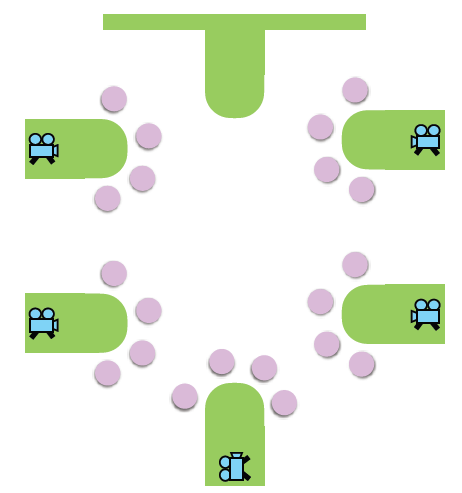}
 \caption{Layout of the IMPRESS classroom. Students sit at the five tables with cameras, the table on the far left is the instructor table.}
\label{fig:room}
\end{figure}

	The context for the present study is the IMPRESS summer program, which is a two-week program for matriculating RIT students who are first generation students and/or Deaf/hard of hearing students (DHH)~\cite{Franklin2017IMPRESS}. This program is designed to serve as a bridge program for students to learn how to reflect on, evaluate, and change their own thinking through intensive laboratory experiments, reflective practices, and discussion both in small groups (3-4 students) and with the whole class (19-20 students). Figure~\ref{fig:room} shows the layout of the room where most of the meetings took place.

    The main objectives of the IMPRESS program are to engage students in authentic science practice, to facilitate the development of a supportive community, and to help the students reflect on the science and themselves in order to strengthen their learning habits and lead them to a stronger future in science, technology, engineering, and mathematics (STEM) fields.

\section{Social Networks}

    Social networks are representations of the connections between groups of people. Our networks are formed of people (nodes) interacting with each other (ties). Ties are directed from the source node (i.e., the initiator of an interaction) to the target node(s) (i.e., receiver(s) of that interaction). In order to perform an analysis of the social networks, we generate graphs that help to visualize each network as well as to study networks quantitatively. 
    
    In this paper, we use video data of the 2016 IMPRESS summer program. 
    We quantify the types of interactions we see in the video to create a social network for the class. 
    Our goal is to track the evolution of conversation topicality within the classroom, as well as overall integration of the classroom network.
    
    In particular, we are interested in characterizing and exploring the developing community of students throughout the duration of the IMPRESS program. We analyze how the patterns of verbal and American Sign Language (ASL) interactions vary over the course of the two-week long program for the class. We distinguish between the topicality of the interactions to help us characterize the development of the students' collaboration.
    
\section{Data Collection}
    
    For the duration of the program, each group of students is recorded by a camera as in Fig.~\ref{fig:room}. We analyzed 30 minutes (per table) of activity time from two different days of the IMPRESS program, day 2 and day 7. On each of these days, students performed  experiments related to climate change. On day 2, they attempted to construct a model of earth or earth's atmosphere using equipment provided to them. Using this model, they then turned on a heat lamp and made temperature measurements in an attempt to capture the effects of global climate change. On day 7, they performed a chemistry experiment. Burning a wooden tongue depressor, they attempted to measure the amount of carbon dioxide (CO$_2$) released. Additionally, we surveyed but did not code data from day 8 of the program. The experiment on this day was measuring the height of the blast from a classic coke and mentos mix, once with room temperature coke, and once with refrigerated coke. On each day, the students construct and execute an experiment while the instructors move from group to group and assist. We should note the instructors neither encourage, nor discourage off-task discussion, and they participate in both on and off-task discussion with the students.
    
    Before the program begins, all students are given pseudonyms  which are then used in research and analysis. Additionally, each student self-reported gender (8 female, 11 male), hearing status (4 students identified as DHH), and other demographic information which is subsequently associated with their pseudonym to allow for demographic based research.
    
    Using the Behavioural Observation Research Interactive Software (BORIS), an event logging software for audio and video analysis ~\cite{BORIS}, we extract network data from videos by identifying spoken and ASL interactions between individuals in a given clip of a fixed length as ties from the source (speaker/signer) to the targets (listeners). We code for three types of interactions: {\bf on-task}, which are specifically  about the given assignment; {\bf on-topic}, which are about STEM and the program more generally; and {\bf off-topic}, which includes all other interactions. The last two classes of interactions align with what Langer-Osuna {\it et al.} call off-task interactions~\cite{Langer-Osuna2018}.
    
    The video data is coded in 100 second intervals (bins). In a bin, each person can send at most one tie of each type to every other person in the network. This grain size is the result of an iterative process within the research group to balance coding effort with information reduction. The grain size for coding was chosen so the network graph for each bin is neither empty nor completely saturated (i.e., with all possible ties). This results in 18 bins of data for each day.
    
    Determining whether an interaction occurred and how it should be classified is a subjective process. To control for this subjectivity, a codebook was created and iteratively refined 
    with the whole research group. In addition, two different researchers coded the same video, discussing differences until they reached a communal understanding. In total, 20\% of the video data was double coded with an inter rater reliability of more than~80\%.

\section{Analysis and Discussion}
	   
\begin{figure}[t]
    \includegraphics[width=.35\textwidth]{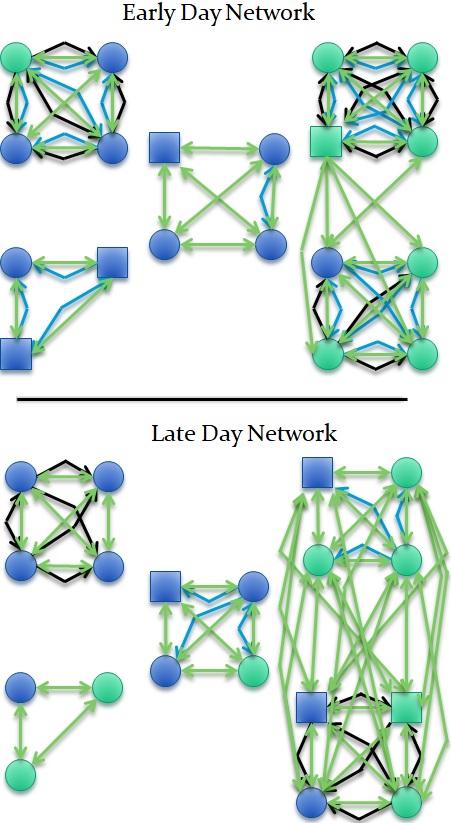}
     \caption{The aggregate networks for the entire room on days 2 and 7. \textit{on-task} ties are in green, \textit{on-topic} ties are in black, and \textit{off-topic} ties are in blue. Square nodes are DHH students (circles are non-DHH), and blue nodes are male identifying students, while green nodes are female.}
    \label{fig:net}
    \end{figure}

    We use the igraph package~\cite{igraph} within the R programming language~\cite{R} to generate graphs representing the recorded data. Figure \ref{fig:net} shows the networks that were generated for the class of 19 students on Days 2 and 7 of the IMPRESS program. We compute the network densities (Day 2: 0.34, Day 7: 0.31) and average path lengths (APL) (Day 2: 14.57, Day 7: 14.49) for both days (a missing path between nodes was counted as 19, that is one more than the maximum possible distance between two nodes in a 19 node). The density and APL are calculated based on fully aggregated and completely flattened networks without instructors.
    These metrics showed no difference in the level of integration of the classroom network.
    
    Additionally, in Fig.~\ref{fig:net}, you can see that particular demographics had no apparent impact on the student's integration in the network. DHH students are as involved in the network as non-DHH students, and gender status similarly made no difference. Overall, our network analysis showed no evidence of an increase in the integration of the overall student community.
    
\begin{figure}[t]
    \includegraphics[width=0.35\textwidth]{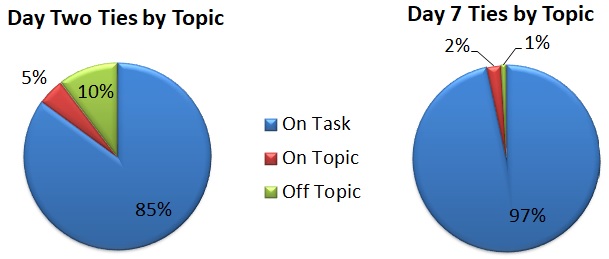}
     \caption{Percentage of ties by topic on each day. Notice the substantial decrease in percentage of off-task ties on day 7 (3\%) as compared to day two (15\%).}
    \label{fig:pie}
    \end{figure}

    We also determine the proportion of interactions of each type on each day. As can be seen in Fig.~\ref{fig:pie}, there is a noticeable change in the topicality of interactions between these two days. On day 2, there is a high proportion of {\it off-task} interactions (15\%). By day 7, almost all interactions (97\%) are {\it on-task}. All of this data is taken from classroom discussion. Programmatically, there are things happening outside of class that are related to community development, but we don't have data from out of class interactions.
    
    As discussed earlier, we explore two possibilities for this variation in topicality. 
    The first possibility is that different assignments drive very different student conversations. To control for this, we looked at 30 minutes of data from day 8. While we did not fully code it, we made observations of the different conversations throughout the activity. We observed that all five groups remained entirely {\it on-task} for the task's duration. Since the activities on days 7 and 8 were completely different and had similar topicality distribution, we believe that it is unlikely that different classroom activities were the driving factor behind conversation topicality shift.
   
    Our data clearly show a decrease in the proportion of off-task interactions. As discussed by Langer-Osuna~{\it et al.}~\cite{Langer-Osuna2018}, productive off-task interactions are primarily a method of positioning. Students use off-task interactions to gain entrance to conversation, bring in students who are not participating, resist students who have commandeered a position of authority, and to gain group investment into a particular idea. In later days of the program, students apparently have less need to resort to off-task interaction to gain support for their ideas, or position themselves within the collaboration.
    
\section{Conclusion}

	Learning takes place not only as a process of memorizing and model construction, but also as participation in a community. In order to maximize student learning, it is important to build strong communities. The IMPRESS program seeks to help students form both good metacognitive practices, and a strong sense of participation in the STEM community. When analyzing student learning and community membership, the types of interactions between students provide vital information.
    
    Our network measures did not show significant changes in network integration. However, early in the program, there was a substantive proportion of {\it off-task} interactions and as the program continued, {\it off-task} interactions vanished almost entirely (during assigned tasks). 
    
    This decrease is consistent with the work of Langer-Osuna~{\it et al.}~\cite{Langer-Osuna2018} where they show that many {\it off-task} interactions serve to position students within a group.
    A decrease in the proportion of {\it off-task} interactions suggests that students have built successful collaborations. Students recognize each other's roles and ideas in the collaboration, reducing the need for off-task interaction. Further research is needed to verify if the overall classroom network integration also increases. Our small group data could not answer this question due to heavily isolated groupwork nature of assignments.
    
\section{Acknowledgements}
This work has been partially funded by NSF grant DUE\#1317450 and by the ``Quality Campaign for Teacher Education'' (QlB) of the German Federal Ministry of Education and Research.

    

%

\end{document}